# High quality-factor mechanical resonators based on WSe$_2$ monolayers

Nicolas Morell[1,*], Antoine Reserbat-Plantey[1,*], Ioannis Tsioutsios[1], Kevin G. Schädler[1], François Dubin[2], Frank H.L. Koppens[1] and Adrian Bachtold[1]

[1]ICFO-Institut de Ciencies Fotoniques, The Barcelona Institute of Science and Technology, 08860 Castelldefels (Barcelona), Spain. [2]INSP, Paris 6 Jussieu. 75006 Paris, France.

[*] These authors contributed equally
Corresponding authors: antoine.reserbat-plantey@icfo.es, adrian.bachtold@icfo.es

**Abstract.** Suspended monolayer transition metal-dichalcogenides (TMD) are membranes that combine ultra-low mass and exceptional optical properties, making them intriguing materials for opto-mechanical applications. However, the low measured quality factor of TMD resonators has been a roadblock so far. Here, we report an ultrasensitive optical readout of monolayer TMD resonators that allows us to reveal their mechanical properties at cryogenic temperatures. We find that the quality factor of monolayer WSe$_2$ resonators greatly increases below room temperature, reaching values as high as $1.6 \cdot 10^4$ at liquid nitrogen temperature and $4.7 \cdot 10^4$ at liquid helium temperature. This surpasses the quality factor of monolayer graphene resonators with similar surface areas. Upon cooling the resonator, the resonant frequency increases significantly due to the thermal contraction of the WSe$_2$ lattice. These measurements allow us to experimentally study the thermal expansion coefficient of WSe$_2$ monolayers for the first time. High $Q$-factors are also found in resonators based on MoS$_2$ and MoSe$_2$ monolayers. The high quality-factor found in this work opens new possibilities for coupling mechanical vibrational states to two-dimensional excitons, valley pseudospins, and single quantum emitters, and for quantum opto-mechanical experiments based on the Casimir interaction.

Monolayer TMDs are two-dimensional direct-bandgap semiconductors that have attracted considerable attention because of their unique optical properties[1-12]. In principle, suspending such monolayer TMDs should form remarkable opto-mechanical resonators. They are extremely thin, like graphene[13-17]. Owing to their low mass, the mechanical vibrational states of TMD resonators are extremely sensitive to external force[18] and adsorbed mass[19]. This holds promise for sensing applications, and for coupling mechanical vibrational states to various optical degrees of freedom of monolayer TMDs, such as bright and dark two-dimensional excitons[1,2], valley pseudospins[3-5], and single quantum emitters embedded in the monolayer[6-9]. However, semiconducting TMD mechanical resonators have been operated only at room temperature thus far[20-25]. Room temperature operation is detrimental to opto-mechanical experiments, since the recombination and coherence time of two-dimensional excitons and valley pseudospins are short, and single quantum emitters only emerge at cryogenic temperatures. An even greater obstacle for all these experiments is the low quality factor $Q$ achieved so far in monolayer TMD mechanical resonators ($Q \leq 100$)[20,22,23]. Therefore, it is important to develop a method to measure TMD resonators at cryogenic temperature in order to reduce damping. Electrical mixing techniques have been used to measure nanotube, graphene resonators and other metallic nanosystems at helium temperature[14-16,26-28], but such techniques are challenging to apply to semiconducting TMD resonators without Joule heating because of their high electrical resistance. Here, we show that optical detection of TMD resonators can be employed down to 3.5 K without being affected by laser heating. This is possible because we have found that the $Q$-factor becomes extremely high at low temperature, allowing us to detect mechanical vibrations with low laser power.

We produce drumhead nano-resonators based on WSe$_2$ monolayers. The fabrication relies on the dry transfer of thin WSe$_2$ crystals over pre-structured holes (Fig. 1a,b)[29]. Regions corresponding to monolayers are identified by the enhanced emission in photoluminescence maps and the corresponding spectra (Fig. 1 c,e)[1,2,10]. At 3.5 K, these spectra feature narrow peaks associated with two-dimensional excitons and trions, labelled as X$^0$ and T, respectively[4]. The observation of trions indicates that the monolayer is doped, which we attribute to molecules adsorbed on its surface.

The mechanical vibrations are detected by optical interferometry[13,30] using ultra-low laser power down to 70 nW in order to prevent heating. A continuous wave laser impinges on the device, and the reflected laser light is modulated by an amount proportional to displacement of the resonator. More specifically, the laser forms a standing wave pattern in the direction perpendicular to the Si substrate, such that the displacement of the WSe$_2$ monolayer modifies its optical absorption[13,30]. In order to prevent heating of the WSe$_2$ lattice, the laser power is kept as low as possible, and the reflected light is collected with a large numerical aperture objective, and detected with an avalanche photodetector (APD). The resulting imprecision of the mechanical displacement is limited by the shot noise of the laser down to about 1 $\mu$W (*cf.* Supplementary Fig. 1). This allows us to detect the driven vibrations of WSe$_2$ monolayers at 3.5 K with laser power down to 70nW. We record the spatial shape of the mechanical mode and its resonant frequency by confocal microscopy (Fig. 1d,f). Motion is actuated by applying an oscillating voltage between the monolayer and the back-gate of the substrate[13] which form together a capacitor. This capacitive actuation is possible because the monolayer is doped. Measurements presented in this Letter are carried out in the linear regime; an example of nonlinear Duffing response is shown in Supplementary Fig. 5.

The dramatic improvement in quality factor by cooling can be seen in Fig. 2a, which shows the frequency spectrum of two driven resonators at different temperatures. For the best resonator, we observe ultra-high quality factors up to $Q = f_m/\Gamma_m \sim 47000$ at helium temperature, where $f_m$ is the resonant frequency and $\Gamma_m$ the resonance line-width. This is more than two orders of magnitude higher than the $Q$-factor of monolayer TMD resonators previously measured[20,22,23]. This also surpasses by a factor 3 the highest measured $Q$-factor of graphene resonators with similar dimensions and cooled at helium-4 temperature[14,15]. Upon increasing the cryostat temperature, the $Q$-factor decreases (Fig. 2b). It reaches $Q \sim 166$ at room temperature, a value consistent with previous reports[20,22,23].

Central to achieving these ultra-high $Q$-factors is the low laser power used to detect the vibrations. Figure 3a displays the quality factor as a function of laser power $P$. We find that the quality factor remains approximately constant for $P$ below $\sim 200$ nW. For larger $P$, the $Q$-factor decreases, indicating that the laser perturbs the dynamics of the resonator. In the following, we show that the laser affects the resonator through two physical processes: absorption heating and dynamical photothermal back-action[30,31]. In absorption heating, the laser increases the temperature of the device lattice, so that the damping $Q^{-1}$ is expected to increase (as shown in Fig. 2b). Dynamical photothermal back-action arises from the gradient of the photothermal force and the finite response time of the resonator to a temperature change[31]. The resulting retarded force modifies the damping of the resonator by $\Delta(Q^{-1})$ and, therefore, its effective temperature. Depending on the sign of the gradient, $\Delta(Q^{-1})$ is positive and the resonator cools down, or $\Delta(Q^{-1})$ is negative and the resonator heats up[31].

We measure the temperature of the mechanical mode by detecting the thermal vibrations of the resonator (Fig. 3c). The capacitive force is switched off, and the resulting displacement noise is recorded with a spectrum analyser. Figure 3c shows that upon increasing $P$, the integrated area of the thermal resonance and its linewidth get both larger with the cryostat temperature set at $T_{cryo} = 20$ K. This shows that the temperature $T_{mode}$ of the mechanical mode and the damping $Q^{-1}$ both increase, in agreement with the absorption heating process.

The thermal resonance area $\langle z^2 \rangle$ in the displacement spectrum allows to quantify $T_{mode}$ using $m_{eff}\omega_m^2\langle z^2 \rangle = k_B T_{mode}$ with $m_{eff}$ the effective mass of the mechanical mode and $\omega_m = 2\pi f_m$. Figure 3d shows that $T_{mode}$ scales roughly linearly with $T_{cryo}$ at high temperature, indicating that the resonator thermalizes with the cryostat. However, the resonance area at $T_{cryo} = 20$ K corresponds to $T_{mode} \sim 50$ K, showing that absorption heating is substantial. The reason for this observed heating is the particularly large laser power needed to resolve thermal vibrations, i.e. $P > 10$ μW.

We find that dynamical photothermal back-action also lowers the $Q$-factor. For temperature cryostat above 70 K, we find that the mechanical mode gets effectively colder upon increasing $P$ (*cf.* Supplementary Fig. 2), indicating that dynamical photothermal cooling becomes more efficient than absorption heating. The dynamical photothermal cooling is related to the force gradient $\partial F/\partial z$ induced by the standing wave of the laser in the direction perpendicular to the substrate, as shown in Ref.[30]. The cooling of the mode observed in our experiment indicates that $\partial F/\partial z > 0$ and $\Delta(Q^{-1}) > 0$ [31]. This confirms that the high $Q$-factors measured at low power in Fig. 2 do not originate from the narrowing of the mechanical linewidth due to dynamical photothermal back-action.

The high $Q$-factors of capacitively driven vibrations measured at low laser power are obtained upon tuning the back-gate voltage $V_g^{dc}$ as close as possible to the equivalent work function difference $\Delta\phi$ between the monolayer and the back-gate (Fig. 3b). Otherwise, the $Q$-factor can significantly decrease due to the electronic Joule dissipation of the displacement current through the monolayer, the current being generated by the TMD motion. The contribution of the Joule dissipation is

$$Q_J^{-1} = R \frac{{C_g'}^2 (V_g^{dc}-\Delta\phi)^2}{m_{eff}\omega_m} \qquad (1)$$

where $C_g'$ is the derivative of the capacitance between the suspended monolayer and the back-gate with respect to the displacement, and $R$ is an effective electrical resistance of the monolayer[27]. Our data are well described by $Q_{total}^{-1} = Q_{WSe_2}^{-1} + Q_J^{-1}$ where $Q_{WSe_2}$ is the quality factor measured at $V_g^{dc} \approx \Delta\phi$ (Fig. 3b). The comparison between the measurements and Eq. 1 leads to $R = 305$ kΩ. Such a value is comparable to the sheet resistance of a TMD monolayer with weak doping[32,33]. We emphasize that a direct comparison between $R$ and the sheet resistance is nontrivial, because of the uncertainties in the geometry associated with $R$.

We now turn our attention to the temperature dependence of the resonant frequency, which is found to be particularly strong (Fig. 4a). This behaviour is attributed to the thermal expansion of the membrane, which modifies the tensile strain $\epsilon$ of the suspended monolayer. The presence of tensile strain is further supported by the convex parabola observed in the dependence of the resonant frequency on the static back-gate voltage (Fig. 4c). The convex parabola, which has been observed in highly strained nanotube and graphene resonators[16], has an electrostatic origin. For a circular membrane under tensile stress, the resonant frequency is given by[34]

$$f_m = \frac{1}{2\pi}\sqrt{\frac{4.92 E_{2D}}{m_{eff}}\epsilon - \frac{0.271\,\epsilon_0 \pi r^2}{m_{eff}\,d^3}\left(V_g^{dc} - \Delta\phi\right)^2}. \qquad (2)$$

Here, $E_{2D} = 116$ N.m$^{-1}$ is the two-dimensional Young modulus of WSe$_2$ monolayer[35], $r = 1.5$ μm is the drum radius, and $d = 180$ nm is the equivalent separation between the membrane and the back-gate electrode. The measurements can be well described by Eq. 2 using $m_{eff} = 3 \cdot 10^{-17}$ kg. The obtained mass is close to the value expected from the two-dimensional mass density $\rho = 12$ μg.m$^{-2}$ of WSe$_2$, which gives $m_{eff} = 0.27 \cdot \pi r^2 \rho = 2.3 \cdot 10^{-17}$ kg. This good agreement shows that the resonator mass predominantly consists of the WSe$_2$ lattice with only a small fraction of additional mass from adsorbed molecules.

Our measurements allow us to study the thermal expansion coefficient $\alpha_{WSe_2}$, an important mechanical property of monolayer TMDs that has not been experimentally investigated thus far. That is, we use the thermal expansion coefficient $\alpha_{WSe_2}(T)$ calculated in the theory work in Ref.[35] in order to reproduce the measured temperature dependence of the resonant frequency. Figure 4b shows that the calculated $\alpha_{WSe_2}(T)$ is positive, in stark contrast to the negative coefficient encountered in many two-dimensional layers, such as graphene[15,36,37] and hexagonal boron nitride[38]. The description of the measured $f_m(T)$ with the calculated $\alpha_{WSe_2}(T)$ is obtained using Eq. 2 with $V_g^{dc} \approx \Delta\phi$ and $\epsilon = \epsilon_0 - \int_{300\,K}^{T} \alpha_{WSe_2}(T_0)\,dT_0$ and using the built-in strain $\epsilon_0$ at room temperature as a fitting parameter (Fig. 4a); the best agreement between the measurements and Eq. 2 is obtained with $\epsilon_0 = 0.034\%$. The thermal expansion of the substrate contributes to $\epsilon$ by a small amount since $\alpha_{Si}(T), \alpha_{SiO_2}(T) \ll \alpha_{WSe_2}(T)$; an upper bound of the substrate contribution to the resonant frequency is indicated by the grey area in Fig. 4a. The temperature dependence of the resonant frequency of the other measured WSe$_2$ resonators is similar to that presented here (*cf.* Supplementary Fig. 3), which further confirms the thermal expansion coefficient in Fig. 4b. Overall, our measurements show that the thermal expansion coefficient of WSe$_2$ is positive and reaches a high value at room temperature ($\alpha_{WSe_2} \sim 7 \cdot 10^{-6}\,K^{-1}$). The precise knowledge of the thermal coefficient of WSe$_2$ over a wide temperature range is important for heat management and designing future (opto-)electronic Van der Waals heterostructures.

High $Q$-factors are also observed in resonators based on other TMD monolayers, such as MoS$_2$ and MoSe$_2$. Figure 5 shows the $Q$-factor of all the seven TMD resonators measured at helium temperature with P<10µW. The corresponding spectra are shown in Supplementary Fig. 4. All the observed $Q$-factors are at least one order of magnitude higher than the values reported previously and measured at room temperature.

The origin of the high $Q$-factors achieved in this work can be attributed to a combination of different factors. First, the low amount of adsorbed molecules is beneficial, since such contamination can increase dissipation through two-level systems[39,40] and the diffusion of adsorbed molecules[41]. Second, the phonon spectrum of TMDs is expected to result in higher $Q$-factors compared to graphene, because of the lower phonon-phonon scattering rate[42]. Third, the strain of the suspended monolayer can lead to the enhanced $Q$, since the $Q$-factor and the resonant frequency vary strongly in the same temperature range, that is, between ~80 and 300 K. The quality factor $Q = f_m/\Gamma_m$ is indeed expected to get larger when $f_m$ increases, assuming that $\Gamma_m$ remains constant[43–45]. However, the increase of $f_m$ by a factor 2 can only explain part of the ~100 enhancement of the $Q$-factor. The strong temperature dependence of the $Q$-factor can be associated to the spectral broadening that arises from the interplay of the amplitude fluctuations of thermal vibrations and nonlinearities, such as the Duffing nonlinearity and mode-mode coupling[46–48]. Such spectral broadening is expected to reduce for larger strain, as is observed in our experiments. The effect of spectral broadening can be tested with ring-down experiments[23], but measuring the ring-down in our devices requires large laser power, such that the mechanical linewidth becomes limited by dynamical photothermal damping. Further work is needed to clarify the underlying physics governing our observed temperature dependence of the $Q$-factor.

We have reported an ultrasensitive optical readout of monolayer TMD resonators that allows us to reveal their mechanical properties at cryogenic temperatures. We report $Q$-factors up to $4.7 \cdot 10^4$ at 3.5 K, thus surpassing the $Q$-factor of graphene nanoresonators with similar areas. High $Q$-factors are obtained using low laser power in order to prevent absorption heating, and by tuning the back-gate voltage so that the damping contribution due to the electron-vibration coupling is negligible. Upon cooling the resonators, the resonant frequency increases steadily, which is due to the thermal contraction of the monolayer crystal. Because of the combination of the high mechanical $Q$-factor and its fascinating optical properties, TMD monolayer provides a unique platform for nano-optomechanics. We envision hybrid optomechanics experiments in which mechanical vibrations are coupled to single quantum emitters[6–9] embedded in the crystal via the strain field[49–52], and to two-dimensional excitons[1,2] and their valley pseudospins[3–5] with a gradient of magnetic field[11,12]. The mechanical vibrations of TMD monolayers can also be coupled to quantum emitters via the Casimir interaction[53]. When placing a quantum emitter, such as a dibenzoterrylene (DBT) molecule, a few tens of nanometres away from the monolayer, vacuum fluctuations lead to extremely large dispersive couplings, opening new possibilities for quantum opto-mechanics experiments[16].

**Methods: Sample preparation**
We pattern electrodes onto the surface of a 285 nm thick $SiO_2$ layer thermally grown on a p-doped Si chip by electron beam lithography (EBL) and thermal evaporation of Au. Holes of 125 nm depth and with diameters ranging from 2-5 μm are structured using EBL and subsequent reactive ion etching. Trenches are etched between adjacent holes to avoid buckling of the subsequently transferred $WSe_2$ membrane when the device is placed in vacuum. $WSe_2$ is exfoliated mechanically using commercial PDMS sheets. Large $WSe_2$ monolayers on the PDMS are identified under an optical microscope by contrast measurement. Identified monolayers are then transferred to target hole arrays using a three-axis micromanipulator. The result is an array of electrically contacted single layer $WSe_2$ membranes (Fig. 1b).

**Methods: capacitive device actuation and optical readout**
Measurements are performed in a vacuum cryostat from 3 to 300K. Our devices are actuated capacitively using a waveform generator and a low-noise voltage source to provide AC and DC voltage signals. We use a custom-built confocal microscope to locally illuminate the device with 633 nm (HeNe) laser light and read out the reflected light using a high-frequency avalanche photodetector (APD).

**Acknowledgments**
The authors thank M. Lundeberg, and C. Sevik for discussions. AB acknowledges supports by the ERC starting grant 279278 (CarbonNEMS), and the Spanish MEC (MAT2012-31338, FIS2015-69831) associated to FEDER. FK acknowledges support by the ERC Career integration grant 294056 (GRANOP), the ERC starting grant 307806 (CarbonLight), and project GRASP (FP7-ICT-2013-613024-GRASP). We acknowledge support by the EE Graphene Flagship (contract no. 604391), the Severo Ochoa Excellence Grant, and Fundació Privada Cellex.

**Supporting Information**
Response of photodetector as a function of laser power. Effective mode temperature as a function of laser power. Additional data for mechanical frequency as a function of temperature. Additional material for high Q-factor MoS$_2$, WSe$_2$ and MoSe$_2$ mechanical resonators. Non-linear mechanical response of WSe$_2$ resonator.

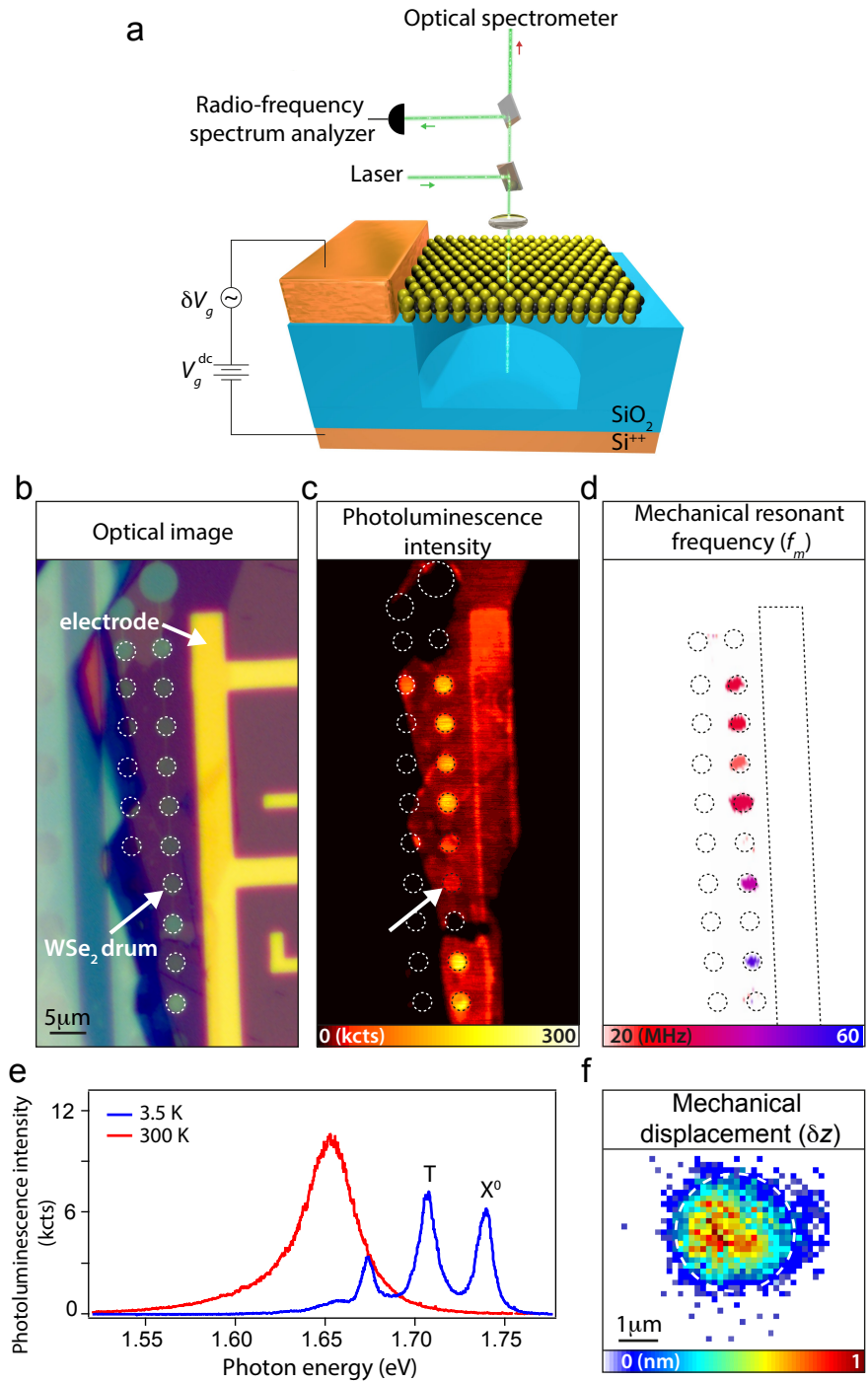

**Figure 1 : WSe₂ monolayer mechanical resonator. a** : Sketch of the device and experimental setup. The WSe₂ resonator is driven capacitively by DC and AC voltages $V_g^{dc}$ and $\delta V_g$, while its nano-motion is measured by optical interferometry. The WSe₂ resonator is a mobile absorber in an optical standing wave produced by a 633 nm probe laser. The modulated laser reflection intensity is then measured with an APD feeding a spectrum analyser. Additional photoluminescence spectroscopy measurements can be done simultaneously using a spectrometer. **b**: Optical image of an array of WSe₂ mechanical resonators. For the same device, both emission and a mechanical spectrum are recorded at each laser position, thus providing a spatial map of the WSe₂ emission (**c**), and the extracted mechanical resonance frequency $f_m$ (**d**). For each position of the laser, we record the mechanical spectrum and we extract the resonance frequency of the fundamental mode. **e**: Photoluminescence spectra of monolayer WSe₂ at 300 K (red) and 3.5 K (blue). **f**: Spatial map of the nano-motion amplitude for a WSe₂ monolayer resonator with the drive amplitude **$\delta V_g = 1mV$**. Dash line represents the WSe₂ resonator outline. The drum studied in (**e,f**) is marked by an arrow in (**b,c**).

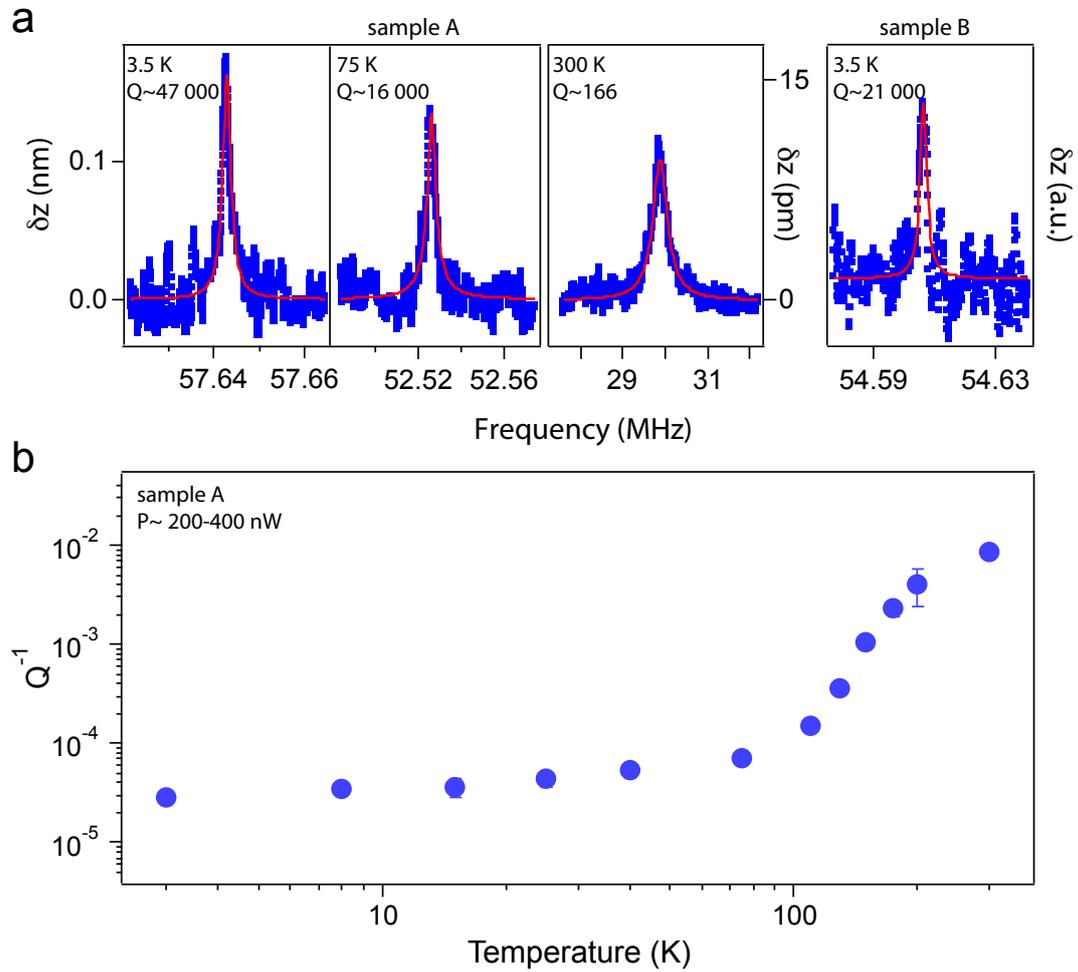

**Figure 2 : High *Q*-factor WSe₂ mechanical resonators. a:** Resonator displacement as a function of drive frequency for devices A (shown in Figs. 1e,f) and B. The solid red line is a lorentzian fit to the data (blue points). We set $\delta V_g$ = 15mV at 3.5K and 75K, and 100 mV at 300K. **b:** Mechanical damping ($Q^{-1}$) as a function of cryostat temperature. The data points below 110 K are taken with 200 nW laser power, and 400 nW otherwise. We use $\delta V_g$ between $10 - 50$mV depending on the temperature.

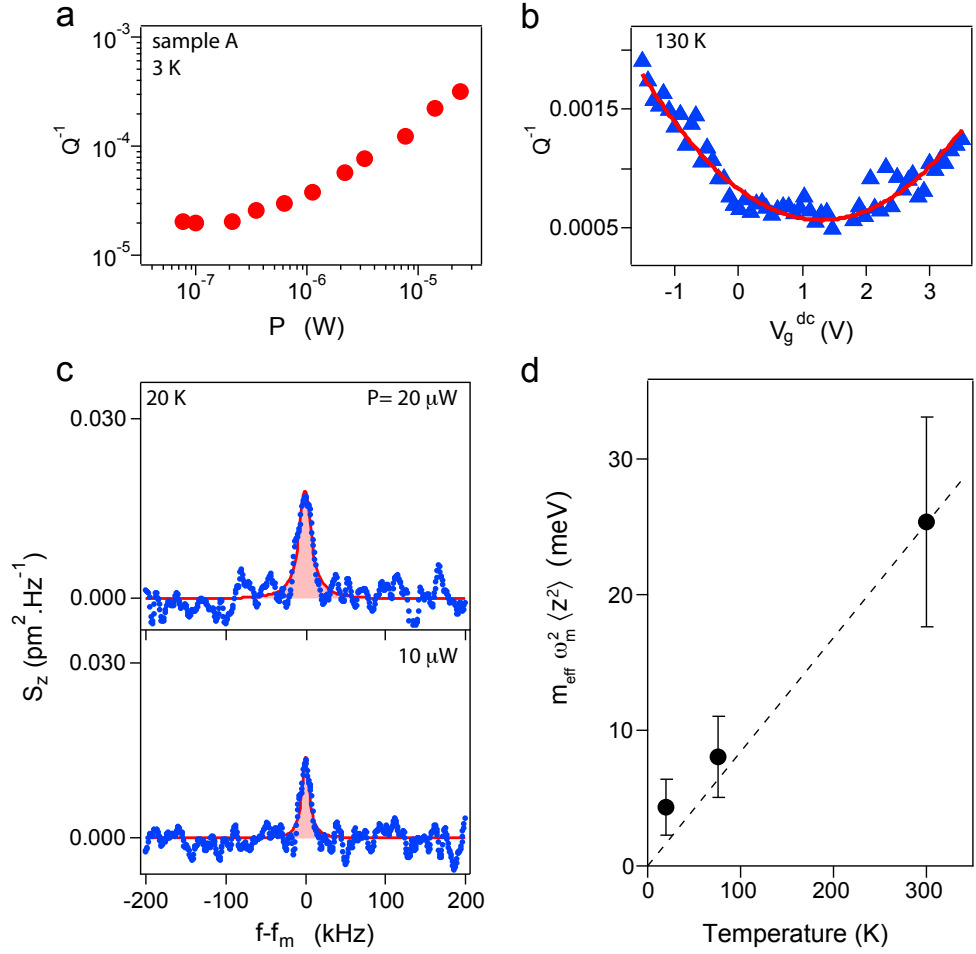

**Figure 3 : Mechanical damping, laser-induced heating, and Joule heating.** Mechanical damping ($Q^{-1}$) as a function of laser power (**a**) and backgate voltage $V_g^{dc}$ (**b**). The red line in (b) is a fit to Eq. 1. We use $\delta V_g = 10mV$. **c:** Resonator displacement noise power density $S_z$ taken at 20K for different laser powers $P$. The area of the Lorentzian fit (red curve) corresponds to the displacement variance $\langle z^2 \rangle$. The increase of this area with $P$ is a signature of laser induced heating. **d**: Energy of the resonator mode $m_{eff}\omega_m^2\langle z^2 \rangle$ extracted from noise spectra as a function of the cryostat temperature. Dashed line corresponds to $m_{eff}\omega_m^2\langle z^2 \rangle = k_B T$.

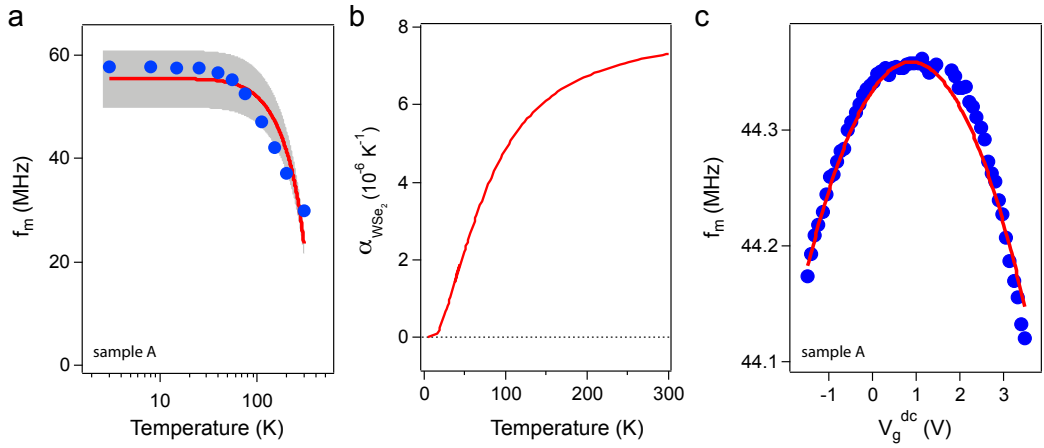

**Figure 4: Resonant frequency of the WSe$_2$ mechanical resonator.** **a:** Mechanical resonant frequency $f_m$ as a function of temperature. We use $\delta V_g$ between $10 - 50\text{mV}$ depending on the temperature. Red curve in (**a**) is a fit to Eq. 2 with $V_g^{dc} \approx \Delta\phi$ and using the thermal expansion coefficient $\alpha$ of WSe$_2$ predicted in Ref.[35] and shown in (**b**). The grey area represents the imprecision due to the thermal expansion coefficient of the substrate. **c:** $f_m$ as a function of the backgate voltage measured at 130 K. The red curve is the capacitive softening quantified by Eq. 2. We set $\delta V_g = 10\text{mV}$.

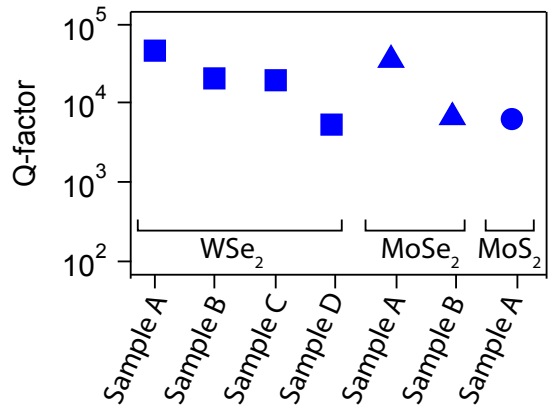

**Figure 5: Quality factors of various resonators based on WSe$_2$, MoSe$_2$ and MoS$_2$ monolayers.** The measurements are carried out at helium temperature with P<10 μW.

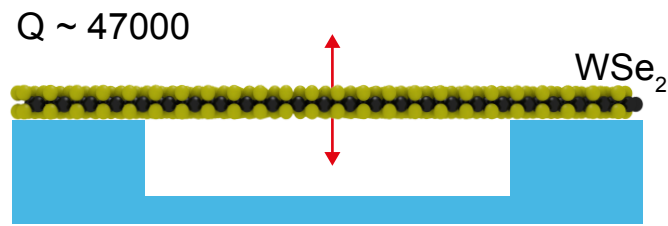

**TOC graphic**